\documentclass[letterpaper, journal]{IEEEtran}
\IEEEoverridecommandlockouts
\usepackage{cite}
\usepackage{amsmath,amssymb,amsfonts}
\usepackage{comment}
\usepackage{algorithmic}
\usepackage{graphicx}
\usepackage{textcomp}
\usepackage{xcolor}
\usepackage[pscoord]{eso-pic}
\usepackage{fancyvrb}
\usepackage{algorithm}

\def\BibTeX{{\rm B\kern-.05em{\sc i\kern-.025em b}\kern-.08em
    T\kern-.1667em\lower.7ex\hbox{E}\kern-.125emX}}

\def\BibTeX{{\rm B\kern-.05em{\sc i\kern-.025em b}\kern-.08em
    T\kern-.1667em\lower.7ex\hbox{E}\kern-.125emX}}
\def\BibTeX{{\rm B\kern-.05em{\sc i\kern-.025em b}\kern-.08em
    T\kern-.1667em\lower.7ex\hbox{E}\kern-.125emX}}

\usepackage{listings}

\usepackage{tabularx}
\usepackage{pifont}
\usepackage{framed}
\usepackage{soul}
\usepackage{booktabs}  
\usepackage{csquotes}
\usepackage{multirow}
\usepackage{array}
\newcolumntype{L}[1]{>{\raggedright\let\newline\\\arraybackslash\hspace{0pt}}m{#1}}
\newcolumntype{C}[1]{>{\centering\let\newline\\\arraybackslash\hspace{0pt}}m{#1}}
\newcolumntype{R}[1]{>{\raggedleft\let\newline\\\arraybackslash\hspace{0pt}}m{#1}}
\usepackage{url}
\MakeOuterQuote{"}
\usepackage[caption=false,font=footnotesize,labelformat=simple]{subfig} 

\usepackage[bookmarks=false]{hyperref}

\algsetup{linenosize=\small}
\usepackage{adjustbox}

\definecolor{mygreen}{rgb}{0,0.6,0}
\definecolor{mygray}{rgb}{0.5,0.5,0.5}
\definecolor{mymauve}{rgb}{0.58,0,0.82}

\makeatletter
\newcommand\notsotiny{\@setfontsize\notsotiny\@vipt\@viipt}
\makeatother

\begin{document}
\bstctlcite{IEEEexample:BSTcontrol}

\newcommand{\toolname}{CWEEP}

\title{%
    CWEEP: A Lexical Static Analysis Framework for CWE Early Prevention
}


\author{Bryan Kwan and Benjamin Tan~\IEEEmembership{Senior Member, IEEE}%
\thanks{B. Kwan and B. Tan are with the Department of Electrical and Software Engineering, University of Calgary, Alberta, Canada. Email: \{bryan.kwan,benjamin.tan1\}@ucalgary.ca}
}


\maketitle

\begin{abstract}
As the hardware layer becomes a focus point for attackers, the need for improved hardware security verification techniques is more important than ever. State-of-the-art security verification techniques require significant manual effort from individuals with security expertise. Furthermore, there is no standard method to locate where the fault lies within the register transfer level (RTL) code. This paper presents \toolname, a static analysis framework for detecting security weaknesses in RTL. \toolname{} does not require a detailed security specification, so it can be used in the early stages of RTL development while properties are still under construction. Furthermore, \toolname{} can identify the exact location in the RTL where the potential vulnerability resides and supports automatic code repair suggestions when applicable. Using datasets from the literature, we evaluate the performance of \toolname{} on a set of two SoC designs with manually inserted bugs and on a large language model generated dataset, consisting of 3874 buggy modules. We find that \toolname{} issues a correct warning up to 60.8\% of the time. In contrast, the tool from a previous work issued a correct warning 17.5\% of the time for the same dataset.
\end{abstract}

\section{Introduction\label{sec:Introduction}}

\begin{figure}[t]
    \centering
    \includegraphics[width=1\linewidth]{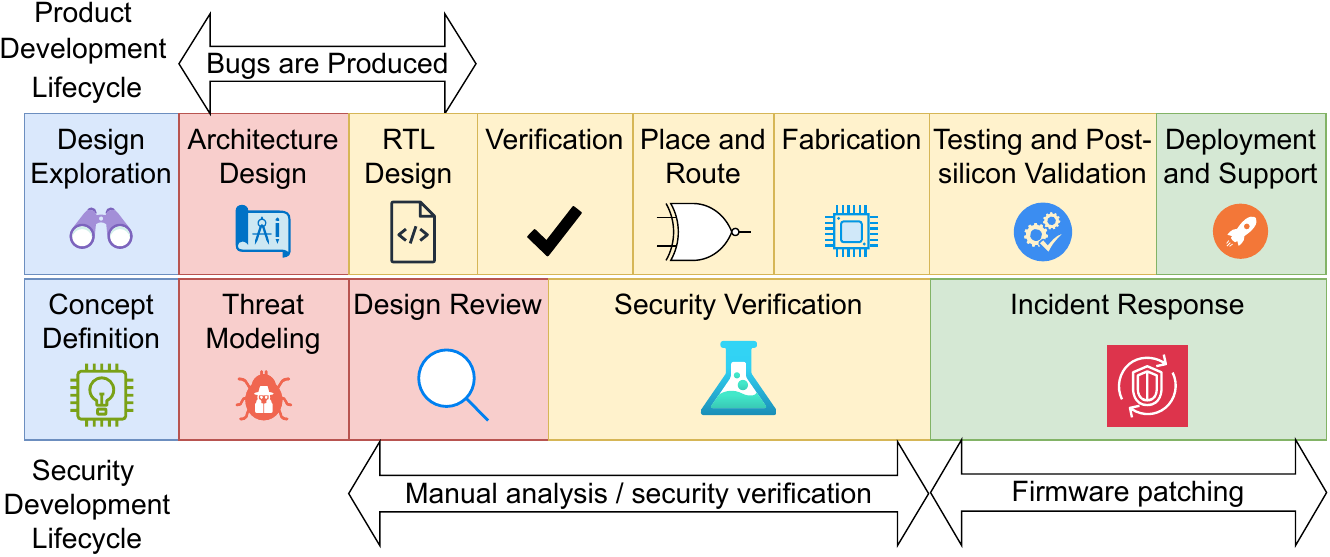}
    \caption{\small Security Development Lifecycle (SDL) and Product Development Lifecycle (PDL) adapted from \cite{dessouky_hardfails:_2019}}
    \label{fig:SDL}
\end{figure}

Growing awareness of security issues at the hardware level~\cite{dessouky_hardfails:_2019,kocher_spectre_2019,lipp_meltdown_2018} motivates the need for techniques to evaluate and improve security, especially in earlier stages of the design flow.  
Prior work~\cite{khattri_hsdl_2012} proposes the Security Development Lifecycle (SDL), which is typically conducted in parallel with the digital product development cycle \cite{dessouky_hardfails:_2019}, as shown in \autoref{fig:SDL}. Existing methods for the various tasks, such as design review, often require extensive manual efforts and expertise. Manual inspection of the design is tedious and error-prone. Moreover, tooling for security verification often requires an extensive and rigorous definition of the security specifications. Considering the intertwined nature of the SDL and the Product Development Cycle (PDL), it would stand to reason that finding security bugs earlier in the cycle is more beneficial. The so-called rule of ten \cite{bushnell_essentials_2004} implies that the cost of fixing a bug found in each stage of development multiplies by approximately ten as we progress through the cycle. Although hardware-level patching techniques have been proposed, such as those in \cite{liu_hardware-supported_2023}, this is not typical in most designs (e.g. \cite{noauthor_lowriscariane_2025}). For this reason, we are motivated to explore security verification techniques that can be used in the earlier stages of development. 

There exist several state-of-the-art techniques for security verification, such as information flow tracking, assertion and property verification, fuzzing, Concolic testing, LLM-based detection, and static analysis. Of these, we explore static analysis as a method for automated/semi-automated bug detection in the earliest stages of RTL design. Prior work \cite{bidmeshki_hunting_2021} identified static analysis tools as one of the first stages in bug finding. This helps enable a \textit{shift-left} approach \cite{noauthor_what_2023}, emphasizing security testing in the earliest stages of development.  

MITRE identifies several vulnerability categories, common weakness enumerations (CWE), which we can use to guide our detection efforts \cite{noauthor_cwe_nodate}. 
Although not all-encompassing, CWE entries 
are often accompanied by offending code snippets or actual examples. 
The CWE database provides a convenient nomenclature for communicating hardware vulnerabilities, and is the chosen framework for a significant portion of security verification work \cite{tarek_bugwhisperer_2025,ahmad_dont_2022,tarek_socurellm_2025,saha_llm_2024,hossain_socfuzzer_2023}.

\textbf{Goals and Contributions}. In this work, we present \toolname, a static analysis framework for CWE-Early-Prevention, improving and expanding on prior related work~\cite{ahmad_dont_2022}. 
We provide a \textit{new} set of algorithms for a selection of CWEs, emphasizing those identified by MITRE as some of the most important weaknesses in industry \cite{noauthor_2025_2025}.
Moreover, we re-implement the algorithms from previous work and provide optimizations for select scanners. 

\toolname{} \textit{also} provides a new solution for bug localization and automatic code repair, highlighting the exact location of the detected vulnerability in the RTL code and, when possible, providing bug fix suggestions. We evaluate the performance of our tool using two HACK@EVENT \cite{dessouky_hardfails:_2019} SoCs. Furthermore, we propose a new evaluation methodology that separates the performance metrics of the two key processes in \toolname. Note that \toolname{} does not obviate the need for human review, nor does it intend to. Instead, we seek to make early-stage hardware security analysis easier by providing a curated list of potential vulnerabilities, reducing the manual burden. 
In summary, our work:
\begin{itemize}
    \item Expands the number of search algorithms compared to prior work~\cite{ahmad_dont_2022}, focusing on key vulnerabilities identified in \cite{noauthor_2025_2025} (\autoref{sec:Proposed Approach}).
    \item Optimizes select algorithms from \cite{ahmad_dont_2022} (\autoref{sec:Proposed Approach}), providing, in the best case, an \textbf{improvement} in precision from 0.167 to 0.778. 
    \item Provides a novel solution that enables bug localization and automatic code repair, which was not present in the prior work by Ahmad \textit{et al.}~\cite{ahmad_dont_2022} (\autoref{sec:Proposed Approach}).
    \item Evaluates our tool, \toolname, using several buggy datasets from the literature, manually analyzing the correctness of each warning raised (\autoref{sec:Experimental Work}).
    \item Introduces a new evaluation methodology that provides additional insight on the mechanism of correct overall detection (\autoref{sec:Analysis Methodology}). 
    \item Provides the full results (and tool implementation) to aid reproducibility (Appendix \ref{sec:Artefacts}). 
\end{itemize}
The rest of this manuscript is as follows. \autoref{sec:Background} positions our work in the literature. \autoref{sec:Proposed Approach} outlines our proposed approach, which we experimentally validate in \autoref{sec:Experimental Work}. \autoref{sec:Discussion} provides additional insights into the feasibility of our proposed approach, and \autoref{sec:Conclusion} concludes. 

\section{Background and Related Work\label{sec:Background}}
\subsection{Related Work}\label{sec:Related Work}
Researchers have focused significant effort on the area of security verification using a myriad of techniques, including information flow tracking \cite{ardeshiricham_register_2017,tai_multi-flow_2021,zapata_automated_2024}, assertion-based property verification \cite{farzana_soc_2019} \cite{ayalasomayajula_lasp_2024}, fuzzing \cite{gubbi_state_2025} \cite{hossain_socfuzzer_2023}, Concolic testing \cite{meng_rtl-contest_2022}, LLM-based bug detection \cite{saha_llm_2024} \cite{tarek_socurellm_2025}, and static analysis \cite{ahmad_dont_2022}. 
In addition to the above vulnerability detection techniques, there are also techniques for bug localization \cite{miftah_rtl-spec_2024} \cite{ma_wit-hw_2025} and automatic code repair \cite{ahmad_cirfix_2022} \cite{laeufer_rtl-repair_2024}.

\textbf{Information Flow Tracking}: This method relies on keeping track of a set of properties in specific signals \cite{ardeshiricham_register_2017}. It tracks confidentiality -- the property that some secret signal $X$ does not propagate to public signal $Y$ -- as well as integrity -- the property that some non-trusted signal $X$ must not allow unauthorized access to $Y$. Information flow tracking (IFT) is a powerful tool for ensuring that sensitive signals do not leak or allow unintentional behaviour. Some IFT techniques, such as \cite{zhang_hardware_2015}, utilize new languages. Although works such as RTLIFT \cite{ardeshiricham_register_2017} extend IFT to operate directly on the RTL, this method still requires manual definitions of design-specific properties, consequently demanding considerable time and expertise. 

\textbf{Assertion-based Property Verification}: As the name suggests, this technique relies on assertions, which are defined based on the design-specific security policies. Commonly, SystemVerilog assertions (SVA) \cite{noauthor_ieee_2024} are used. The construction of these assertions requires significant manual effort. Consequently, there is a focus on automatically generating assertions \cite{iman_survey_2025}. Some works use algorithms to produce the assertions \cite{malburg_property_2017,germiniani_harm_2022,zhang_transys_2020}, whilst others utilize LLMs \cite{ayalasomayajula_lasp_2024}. However, these generated assertions do not always make sense.

\textbf{Fuzzing}: Fuzzing is a technique originating from software that has recently seen applications in hardware security \cite{gubbi_state_2025}. The aim of the approach is to inject a predefined series of inputs in hopes of eliciting unintended behaviour and revealing security vulnerabilities. However, similar to the previous methods, we must define what we consider as unintended behaviour in design-specific security policies, e.g.~\cite{ammann_dy_2024}. While some implementations use code coverage metrics instead of security properties, e.g.~\cite{laeufer_rfuzz_2018}, code coverage is not necessarily indicative of security coverage~\cite{hossain_socfuzzer_2023}.

\textbf{Concolic Testing}: Concolic testing uses symbolic variable inputs during concrete simulation, hence "conc(rete-symb)olic" testing \cite{meng_rtl-contest_2022}. Indeed, there are general principles that Concolic testing can enforce, for instance, no negative addresses. However, many security properties must be defined with specific signals and design-specific details in mind.

\textbf{LLM-based Bug Detection}: Recently, LLMs have shown promise in areas such as security bug detection, threat modelling, and bug insertion \cite{saha_llm_2024} \cite{tarek_socurellm_2025}. This method is more easily extended to a wide range of CWEs compared to other techniques, and typically benefits from improvements in the LLM itself \cite{saha_llm_2024}. However, LLMs tend to incur increasing cost as the models improve in performance \cite{liagkou_cost_2024}.

\textbf{Static Analysis}: Lexical static analysis (which we will refer to as static analysis from hereon), is a software technique which was extended for RTL in tools such as \cite{Snyder_Verilator}. This method was adapted for hardware security in Ahmad \textit{et al.}'s CWEAT~\cite{ahmad_dont_2022}, a static analysis tool for finding CWEs in RTL. On its own, static analysis does not require design-specific context or security policies, since it searches for general patterns, or heuristics, in the code structure that may constitute a vulnerability. We will refer to this as \textbf{contextless} static analysis, i.e., static analysis without design-specific context. As a result, it is possible to run static analysis in the early stages of RTL development without compilation or simulation, unlike methods such as fuzzing and Concolic testing. However, due to its heuristic nature, this technique often results in a high false positive rate. Furthermore, it is difficult to develop static analysis checkers for certain CWEs which rely on design-specific context.

\textbf{Bug Localization and Automatic Code Repair}: Bug localization is the problem of finding which lines of code in the RTL cause the bug. Automatic code repair, as the name suggests, is concerned with autonomously fixing bugs in the RTL code. There is a significant body of work in bug localization \cite{miftah_rtl-spec_2024,ma_wit-hw_2025} and automatic code repair \cite{ahmad_cirfix_2022,laeufer_rtl-repair_2024}, but it is unclear how to extend them to traditional hardware security verification techniques, since they rely on testbench simulations to determine when incorrect behaviour occurs -- hardware security bugs can have obscure execution paths \cite{gubbi_state_2025}, which could impair the efficacy of these methods.
\cite{ahmad_hardware_2024} uses LLMs to repair security vulnerabilities directly in the RTL code. However, it requires the user to prompt the model with the offending code, the correctly identified security vulnerability, and a suggested template for the fix, i.e., it requires prior detection of the bug.

\subsection{Motivation}\label{sec:Motivation}
In general, according to the rule of ten \cite{bushnell_essentials_2004}, it is much more costly to find bugs in the later stages of development; it is ideal to find such bugs during design. Furthermore, the existing security verification techniques tend to require significant manual effort (e.g.~\cite{farzana_soc_2019}). Hence, we desire a solution that can operate directly on the RTL without requiring significant setup effort and manual input. 

Static analysis and LLM-based bug detection search for patterns in the code that look like potential security bugs. In contrast, the other techniques use security properties to decide what behaviour is considered a security vulnerability. In the context of the SDL, static analysis and LLM-based bug detection are the earliest checks, since they do not require design-specific security policies.

The computational cost of static analysis is relatively lightweight, particularly in comparison with LLM-based methods. Hence, we explore static analysis as a method for early-stage bug detection, so that easily detectable bugs do not propagate to later stages of security verification. 

Currently, static analysis suffers from high false positive rates and still requires manual effort to review and implement fixes. Due to its heuristic nature, static analysis tends to produce a significant number of false positives (see \cite{ahmad_dont_2022}). The cost of each warning is relatively high, since a security expert must review the warning, identify the offending RTL code, and develop a fix. Therefore, we propose \toolname, a static analysis tool with an improved framework that provides an end-to-end flow for security vulnerability detection, supporting bug localization and automatic code repair. Our framework is tailored and optimized to reduce the false positive rate compared to previous work \cite{ahmad_dont_2022}. \toolname{} provides the exact location in the RTL where we detected the potential vulnerability, which enables the tool to provide automatic code repair (auto-fix) suggestions for the user to review. Our aim in these improvements is to reduce the manual burden on the reviewer. Additionally, to improve the completeness of our tool, we expand the number of supported scanner algorithms.

\section{Proposed Approach\label{sec:Proposed Approach}}
\subsection{Vulnerability Classification}\label{sec:Vulnerability Classification}
To better organize hardware security bugs, MITRE describes Common Weakness Enumerations (CWEs) \cite{noauthor_cwe_nodate}, which broadly describe different classes of vulnerabilities. To clarify terminology, this work is concerned with the detection of bug patterns, which may exemplify a CWE but do not necessarily comprehensively represent that CWE. However, for convenience, we refer to our algorithms as searching for CWEs.

The authors in \cite{ahmad_dont_2022} provide a classification of which CWEs are statically analyzable. Since then, CWE entries have been updated, and in this work, we offer an alternative classification. Since we are concerned with detecting code patterns, we start by identifying all CWEs with RTL code examples available, since these examples are invaluable for creating our detection heuristics. The code examples consist mainly of bugs from the HACK@EVENT series \cite{dessouky_hardfails:_2019}.

As shown in \autoref{tab:cwe_classification}, we categorize the CWEs with code examples as amenable to \textbf{contextless} static analysis or non-amenable. Among these CWEs, we emphasize (in \textbf{bold}) those that appear in the 2025 Most Important Hardware Weaknesses \cite{noauthor_2025_2025} list. 

To give examples of our classifications, consider CWE-1209: Failure to Disable Reserved Bits. Determining which bits in a register are meant to be reserved is typically infeasible with just the RTL. Consequently, we classify this CWE as non-amenable to \textbf{contextless} RTL static analysis. In contrast, consider CWE-1243: Sensitive non-volatile information not protected during debug. The code example for this CWE indicates that sensitive information should have a mechanism to zero out the value during debug mode. We can check for this bug pattern without any additional context on what the design is doing. The issue of identifying keys is remedied by the keyword-based allowlist used in \cite{ahmad_dont_2022}, which remains the same for a particular class of designs. Therefore, we classify CWE-1243 as amenable to \textbf{contextless} RTL static analysis. 

With this in mind, we select fourteen CWEs (shown in \textit{italic} in \autoref{tab:cwe_classification}) to implement search algorithms for in our tool -- prioritizing those that appear in the most important weaknesses list, as well as those with relative ease of implementation.

\begin{table}[t]
    \small
    \caption{\small Classification of CWEs (based on version 4.18)} \label{tab:cwe_classification}
    \centering
    \begin{tabular}{|p{0.25\linewidth}|p{0.43\linewidth}|p{0.17\linewidth}|}
    \hline
        \textbf{Classification} & \textbf{CWEs with RTL Examples} & \textbf{\# of CWEs} \\ \hline
        Non-amenable to \textbf{Contextless} RTL Static Analysis & 325, 440, 1209, 1220, 1221, 1224, 1241, 1255, 1281, 1298, 1299, \textbf{1300}, 1311, 1317 & 14 \\ \hline
        Amenable to \textbf{Contextless} RTL Static Analysis & \textbf{226}, \textbf{1191}, \textbf{\textit{1231}}, \textit{1232}, \textbf{\textit{1234}}, \textit{1243}, \textbf{\textit{1244}}, \textit{1245}, 1258, \textbf{\textit{1260}}, \textbf{\textit{1262}}, \textit{1271}, \textit{1276}, \textit{1280}, \textit{1310}, \textit{1326}, 1329, \textbf{1431} & 18 \\ \hline
    \end{tabular}
\end{table}

\subsection{Framework}\label{sec:Framework}
We continue our exploration of static analysis by reimplementing the algorithms from previous work (CWEAT \cite{ahmad_dont_2022}). We chose to implement our tool by extending the open source SystemVerilog linting tool, Verible \cite{chipsallianceverible}. The underlying operation is much the same as the previous work \cite{ahmad_dont_2022}, but deals with the concrete representation of the RTL, so it is straightforward to provide the code location (i.e., file, line number, column number) of any node in the CST. This easily solves the issue of bug localization. Furthermore, it helps in our implementation of automatic code repair.

\begin{figure}
    \centering
    \includegraphics[width=1\linewidth]{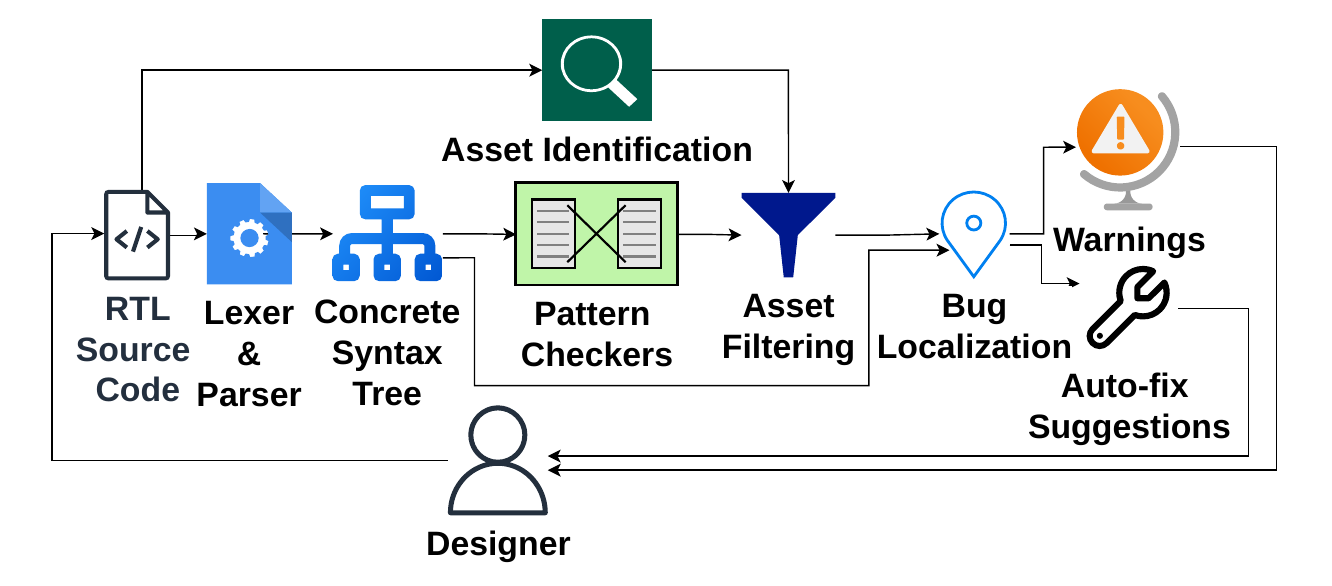}
    \caption{CWEEP: the proposed framework. We first preprocess the source code into a concrete syntax tree. Then, we perform pattern checking on nodes of this tree. Using asset identification, we filter for more relevant matches. During pattern checking, we track the offending node of the concrete syntax tree, which allows us to easily locate the code position of the potential bug. Then, we issue a warning and any relevant auto-fix suggestions to the user.}
    \label{fig:flow}
\end{figure}
\autoref{fig:flow} outlines our proposed flow. We directly extract two things from the RTL, that being the lexical pattern we are searching for, as well as the asset classification of the signal of concern. The pattern checker will search for specific lexical patterns in the code that may constitute a potential weakness. The asset identification method determines whether or not the signal of concern (in the lexical pattern) is an asset relevant to security. We only keep lexical patterns that also contain a signal we consider to be an asset (the Asset Filtering stage of our flow). Then, by carefully tracking relevant nodes within the CST during pattern checking, we locate the offending code. If possible, we also issue an auto-fix suggestion. 

We use the original algorithms of Ahmad \textit{et al.}'s prior work~\cite{ahmad_dont_2022}, as well as our own additions and modifications, to perform the role of pattern checking. Then, we use a keyword-based filtering (implemented via regular expression) to identify assets based on their signal name. Asset identification is not the focus of this work, so we keep the method simple. More complex asset identification algorithms such as \cite{liu_identifying_2025} or \cite{ayalasomayajula_automatic_2024} require significant computational resources, so we opt to not use them in our case, since the intent is to create an early development tool that does not require compilation. However, if performance is the primary concern, substituting our naive asset identification algorithm with one of the aforementioned sophisticated algorithms may prove useful. We note that previous work has shown that keyword filtering can be useful as a minimal way to perform asset identification~\cite{deb_nath_toward_2025}.

Finally, we take advantage of our identified code patterns to provide automatic bug fix suggestions that a user can interactively handle. We again note that this flow is intended not to replace human review of the code, but to augment this process and minimize the manual effort required. 

\subsection{Re-examining Prior Work}\label{sec:Revisiting Prior Work}
Incorporating our basic reimplementation of CWEAT \cite{ahmad_dont_2022} into the framework, we examine the HACK@DAC2021 SoC \cite{noauthor_hack_eventhackatdac21_2024} \cite{dessouky_hardfails:_2019} to get an idea of the noise seen by each scanner algorithm. Our results for which are shown in \autoref{tab:keyword_list_table}, noting the violation counts before and after asset filtering. Our keyword filtering strategy, as outlined later in \autoref{sec:Keyword Filtering}, uses the settings shown in \autoref{tab:regextable}. 

As we do not have access to the source code for \cite{ahmad_dont_2022}, we will primarily use our reimplementation as a benchmark for comparison. Note, however, that we observed our implemention of the scanner for CWE-1271 has a much higher violation count than reported in Ahmad et al.'s work~\cite{ahmad_dont_2022}; the rest of the CWE scanners have a comparable or lower violation count. Since our goal is primarily to minimize false positives, we thus consider our reimplementation to be sufficiently close to the original. In other words, since our reimplementation generally outperforms the original, we conclude that it serves as an acceptable comparison. 

\begin{table*}[t]
\caption{\small Initial Violation Counts}\label{tab:keyword_list_table}
\centering
\resizebox{\textwidth}{!}{
    \begin{tabular}{|c|c|c|}
    \hline
    \textbf{CWE} & \textbf{Violation Count (Before Filtering)} & \textbf{Violation Count (After Filtering)} \\
    \hline
    CWE-1234: Hardware Internal or Debug Modes Allow Override of Locks & 109  & 2   \\ \hline
    		
    CWE-1245: Improper Finite State Machines (FSMs) in Hardware Logic & 165 & 18    \\ \hline
    		
    CWE-1262: Improper Access Control for Register Interface & 12 & 1     \\ \hline
    CWE-1271: Uninitialized Value on Reset for Registers Holding Security Settings & 4853 & 19   \\ \hline
    CWE-1280: Access Control Check Implemented After Asset is Accessed & 394 & 0 \\ \hline
    \end{tabular}
}
\end{table*}

Now we manually analyze, in further detail, the filtered results shown in \autoref{tab:keyword_list_table} to get a sense of how each pattern checker could be improved. The results of our manual analysis can be found in \autoref{fig:hackatdac21analysis} (the methodology for which is explained later in \autoref{sec:Experimental Work}). In our analysis, we find several ways the lexical checkers could be optimized to remove some of the false positives (i.e., the CWE rule is not actually violated due to something the checker missed). 

For CWE-1271, we find that a few of the false positives are raised when the register is connected to a module output. The algorithm does not find a reset and simply raises a violation for that register. However, it does not account for any resets applied to the signal from within the module. Additionally, whenever a register is tied to another signal through an $assign$ statement, the tool raises a violation even if the signal to which it is tied has a reset value.

The algorithm for CWE-1234 checks for a very specific pattern, resulting in only two warnings (one true positive, one false positive), from which we do not draw any patterns.

CWE-1245 sees a comparable violation count, 18 versus the original 17. We find 15 false positives and 3 true positives. The current checker is unable to determine the actual value of each state, which hinders our case completeness check. The state values are defined in a parameter assignment, so it would be beneficial if the algorithm could harvest allowed state values. From our survey of the FSM files we manually analyzed, we find that there are three main ways state information is defined: enumerations, parameter assignments, and define statements. Furthermore, we find that many false positives occur due to unexpected syntax options such as:
\begin{enumerate}
    \item nested case statements, 
    \item comma-separated case values,
    \item the use of ternary operators in case transition statements,
    \item and defining some state transitions inside the case statement, but others outside.
\end{enumerate}
All but the last item are easily managed, since assignments to the state variable outside the case statement may not give a clear beginning and ending state.

For CWE-1262, we see only one violation message, which happens to be a true positive. This matches the one true positive found in CWEAT, but eliminates the 8 other violations (6 false positives and 2 indeterminate).

Interestingly, we see zero violations from CWE-1280, compared to the 12 false positives in CWEAT.

\begin{framed}
\textbf{Takeaway}: Overall, our base reimplementation of CWEAT \cite{ahmad_dont_2022} tends to reduce false positives, allowing us to compare later improvements conservatively (it bounds the original implementation in a beneficial manner). We identify key areas for optimization and improvement in the pattern checkers for CWE-1271 and CWE-1245, which we explore in the following section. We found that the remaining checkers (CWE-1234, CWE-1262, CWE-1280) did not demand optimization. Furthermore, we identified some false positives in this stage, which allowed us to refine our keyword list (which we explain later in \autoref{sec:Keyword Filtering}). For example, we refined our search for locks to block signals such as 'block' or 'pll\_lock'. 
\end{framed}

\subsection{Keyword Filtering}\label{sec:Keyword Filtering} 
As shown in \autoref{tab:keyword_list_table}, the keyword filtering stage is pivotal in narrowing down the number of warnings we need to check. In our simple implementation of the asset identification stage (as shown in \autoref{fig:flow}), we use an allowlist and a blocklist to filter signals by the keywords in their name. For example, we might use the allowlist keyword $lock$, but use the blocklist keyword $clock$. This strategy is used in the previous work \cite{ahmad_dont_2022} and also for a similar application, asset identification, in \cite{deb_nath_toward_2025}. \autoref{tab:regextable} details the allowlist and blocklist settings of each scanner. We constructed these lists by referencing the common keywords in the keyword frequency analysis of \cite{deb_nath_toward_2025} and \cite{ahmad_dont_2022}.

It is important to note that some checkers are only concerned with a specific type of asset (rather than any generic asset), such as CWE-1310 which may look for signals related to the ROM. As such, the keywords for each checker may differ. Additionally, select scanners may need to search for clock or reset keywords. We similarly use the keyword list $clock,\ clk,\ reset,\ rst$ for this purpose. These are not shown in \autoref{tab:regextable} but are mentioned in \autoref{sec:Checker Algorithms}.

\begin{table}[t]
\small
\centering
\caption{\small  Allowlist and Blocklist Settings For Each Algorithm (CWEs without are omitted)}
\label{tab:regextable}
\begin{tabular}{|p{0.08\linewidth}|p{0.25\linewidth}|p{0.5\linewidth}|} 
\hline
\textbf{CWE}  & \textbf{Allowlist}                           & \textbf{Blocklist}                               \\ \hline
1231 & lock, \_lk,reglk & block, clock, arlock, awlock, ar\_lock, aw\_lock, pll, cycle, count   \\ \hline
1232 & lock, \_lk,reglk & block, clock, arlock, awlock, ar\_lock, aw\_lock, pll, cycle, count   \\ \hline
1234 & dbg, debug       &    --                                                           \\ \hline
1243 & key             &  --                                                        \\ \hline
1244 & priv            &   --                                                       \\ \hline
1262 & lock, \_lk, reglk, key, prot, access     & block, clock, arlock, awlock, ar\_lock, aw\_lock, pll, cycle, count            \\ \hline
1271 & lock, \_lk, reglk, key, prot, access     & block, clock, arlock, awlock, ar\_lock, aw\_lock, pll, cycle, count                       \\ \hline
1276 & lock, \_lk, reglk, key, prot, access     & block, clock, arlock, awlock, ar\_lock, aw\_lock, pll, cycle, count                     \\ \hline
1280 & lock, \_lk, reglk, key, prot, access     & block, clock, arlock, awlock, ar\_lock, aw\_lock, pll, cycle, count                       \\ \hline
1310 & rom.*data                           &    --                                                                               \\ \hline
1326 & rom                                 &   --                                                                                \\ \hline
1329 & key                                 &   --                                          \\ \hline                                  
\end{tabular}
\end{table}

\subsection{New and Improved Checker Algorithms}\label{sec:Checker Algorithms}
In this section, we discuss our checker algorithms for a selection of CWEs. We provide entirely new algorithms, as well as optimized versions of algorithms from \cite{ahmad_dont_2022}. We developed each algorithm such that it finds the lexical (syntax) pattern provided in the example of the MITRE definition page \cite{noauthor_cwe_nodate}. Note that for CWE-1234, CWE-1262, and CWE-1280, we repeat them here for convenience, but they are unchanged from the original, given the results we saw in \autoref{sec:Revisiting Prior Work}. 

\textbf{CWE-1231}: Improper Prevention of Lock Bit Modification. 
The CWE-1231 scanner searches for all assignment statements that zero a lock register, which is found via the allowlist. Among these assignments, their parent if statement (if any) must not contain impermissible control signals. At most, one reset is allowed, and any other control signal must have the keyword $lock$. Otherwise, this assignment is flagged as a violation.

\textbf{CWE-1232}: Improper Lock Behaviour After Power State Transition. 
CWE-1232 covers a wide range of scenarios, but we focus on one bug pattern in particular, where lock bits are not set correctly on reset. The algorithm searches all if statement conditions for reset keywords. If any child assignment statements set a lock bit register (found with the allowlist), then we verify that the assigned value is all ones in binary. 

\textbf{CWE-1234}: Hardware Internal or Debug Modes Allow Override of Locks. 
The CWE-1234 checker scans all if statement expressions, searching for the presence of the patterns '$||\ debug$' and '$debug\ ||$'. If either is found, the if statement is flagged as a violation (given that the other operand is an asset). This algorithm is essentially unchanged from \cite{ahmad_dont_2022}.

\textbf{CWE-1243}: Sensitive Non-Volatile Information Not Protected During Debug. 
This scanner checks all $assign$ statements for security-sensitive signals matching the allowlist. For each of these statements, it parses for ternary operators. If there is not a ternary operator with one of its values constant, the statement is flagged as a violation. In essence, we are checking for sensitive information tied to a non-constant value, which could indicate a weakness. 

\textbf{CWE-1244}: Internal Asset Exposed to Unsafe Debug Access Level or State. 
Although CWE-1244 is a broad vulnerability that can manifest in various ways, we provide a scanner to check for a specific RTL bug pattern resulting from debug signal overrides. We check all assignments, where the left-hand side signal matches the allowlist corresponding to privilege signals, for expressions containing '$||\ debug$' and '$debug\ ||$'. Such assignments are flagged as violations. This checker is almost identical to CWE-1234, but checks assignments to privilege signals rather than if conditions. 

\textbf{CWE-1245}: Improper Finite State Machines (FSMs) in Hardware Logic. 
The CWE-1245 scanner is reworded somewhat differently from \cite{ahmad_dont_2022}, but the general steps are similar. The stages of the algorithm can be decomposed into two stages: data collection
and rule checking.

In the data collection stage, we first check all case statements with case variables belonging to an enumeration declaration or matching the allowlist (keyword $state$); any variable names found in an enumeration are added to the allowlist, enabling us to handle state machines with multiple state names, e.g. $state$ and $state\_next$. Then, we attempt to extract any state definitions from enumerations, define statements, and parameter declarations. Any state definition we find initializes the vertices of a state transition graph. Finally, we extract each case item and child assignments involving the state variable to build the edges of the state transition graph. Upon completion of the graph, we compute the indegree and outdegree of each vertex. As suggested earlier in \autoref{sec:Revisiting Prior Work}, we ensure that we are able to parse comma-separated case values and ternary operators; we simply skip nested case statements.

In the rule checking stage, we scan the graph for the following issues, flagging the case statement as a violation if any are present:

\begin{enumerate}
    \item A vertex with indegree 0 is an unreachable state.
    \item A vertex with outdegree 0 is a dead-end state (referred to as a deadlock state in \cite{ahmad_dont_2022}).
\end{enumerate}

Furthermore, we check the case statement for completeness, similar to \cite{ahmad_dont_2022}. If we extracted any state definitions in the data collection stage, we will detect if they are not implemented in the case statement, since the outdegree will be 0. We also check for missing values in the sequence of case values. We first check whether the states are one-hot encoded. If so, we check for missing powers of two. Otherwise, we check for missing incremental values. Any case statement missing a case for a state value must have a default, or it is flagged as a violation.

\begin{figure}[t]
    \centering
    \subfloat[State Graph]{\includegraphics[width=0.5\linewidth]{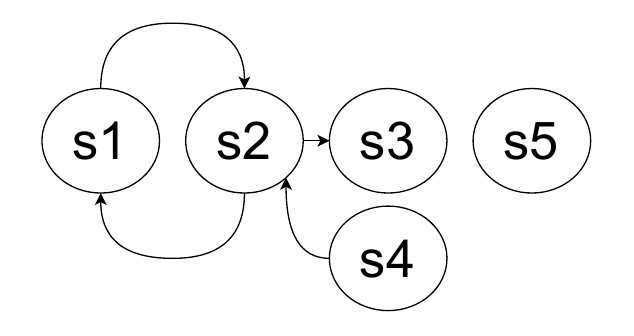}\label{fig:state_transition_example}}
    \subfloat[Indegree and Outdegree]{
        \adjustbox{valign=b}{
            \small
            \begin{tabular}{c|c|c}
            State & Indegree & Outdegree  \\
            \hline
            s1    & 1                            & 1                        \\
            s2    & 2                            & 2                         \\
            s3    & 1                            & 0                         \\
            s4    & 0                            & 1                          \\
            s5    & 0                            & 0                          
            \end{tabular}
        }\label{tab:state_transition_example_indegree_outdegree}
    } 
    \caption{\small State transition example of an FSM with five states. States s3 and s4 are dead-end and unreachable states, respectively. State s5 is not specified in the case statement and is shown as disconnected from the rest of the graph.} 
    
\end{figure}

For example, consider the state graph in \autoref{fig:state_transition_example}. The indegree and outdegree of each node are shown in \autoref{tab:state_transition_example_indegree_outdegree}. Our scanner will identify states s3, s4, and s5, since they have either an indegree or outdegree of 0. States like s5 are typically initialized as a vertex with no edges when we extract the state from declarations like enumerations. If that is the case, we will also flag the case statement for incompleteness if there is no default case. 

\textit{Algorithm Summary}: We build a state graph (by parsing the FSM declaration in the code) and check for unreachable states or dead-end states. Then, we check for missing state definitions by examining either state type declarations or the one-hot encoding of the state signal. 

\textbf{CWE-1260}: Improper Handling of Overlap Between Protected Memory Ranges.  
The implementations of memories can vary greatly, so we demonstrate a scanner that checks for bug patterns in a specific type of memory region definition. We examine the pattern XBase and XLength, i.e., a common identifier, $X$, concatenated with one of the two keywords $base$ and $length$. 

This scanner parses all parameter and enum assignments for matching pairs of signals of the form XBase and XLength. For example, $GPIOBase$ and $GPIOLength$. Then, it searches for constant assigned values in hexadecimal format. Then, it defines an address region based on those two values for the identifier. Comparing all memory regions, we can then detect overlap if the expression $max(base1,base2) <= min(end1,end2)$ evaluates to true, where $base1$ and $base2$ are the base addresses of the two regions, and $end1$ and $end2$ are their respective ending addresses. In the case of overlap, we flag both offending assignments as violations.

\textbf{CWE-1262}: Improper Access Control for Register Interface. 
First, we find all assignments to security-relevant signals (found via keyword list) with the right-hand side of the assignment containing the keyword $wdata$. Then, we extract control signal names from parent if statements or ternary operators (clock and reset keywords are ignored). Registers with no control signals are flagged as violations. Additionally, for register arrays, each member of the array must have the same control signals. Otherwise, we flag a violation. 

\textbf{CWE-1271}: Uninitialized Value on Reset for Registers Holding Security Settings.
Similar to the algorithm provided in \cite{ahmad_dont_2022}, we scan for all register variables matching the allowlist and check that there is a matching reset statement for that register. In \cite{ahmad_dont_2022}, the register is simply flagged as a violation if there is no matching reset statement. 

However, our implementation provides two optimizations to reduce the number of false positives. The first optimization is to prune all registers that are tied to module outputs, which are naively detected by searching for the suffix $\_o$ in the port name. This method could be improved, but it was sufficient for our dataset. The idea is that if there is actually a violation on that signal, we will flag it when scanning the file containing the actual module definition. The second optimization is to follow $assign$ statements. For example, if our register signal is tied by $assign$ statement to another signal, we check the rule for CWE-1271 on that intermediate signal instead. 

\textbf{CWE-1276}: Hardware Child Block Incorrectly Connected to Parent System.
This vulnerability can materialize in a plethora of ways. Hence, we focus on one specific case that is, in general, a good idea to fix: floating port connections. This feature is often available in standard linting tools, but we filter the results only to port names matching the security keyword allowlist, in hopes that this will help security-focused analysis. The checker simply searches for all port connections with names matching the allowlist, and raises a violation if the connection is empty (floating).

\textbf{CWE-1280}: Access Control Check Implemented After Asset is Accessed. 
Our checker searches for always blocks and iterates through every child blocking assignment in sequential order. We track the history of all signal accesses, i.e., signals on the right-hand side of the assignment, and ensure that those signals are not assigned afterwards. Otherwise, we flag both assignments as violations. 

\textbf{CWE-1310}: Missing Ability to Patch ROM Code. 
We search all assignments to ROM-like variables (filtered via the allowlist). Subsequently, we track all possible values to those ROM-like variables. For each variable, we ensure that it has at least two possible unique values. Otherwise, we flag a violation. 

\textbf{CWE-1326}: Missing Immutable Root of Trust in Hardware. 
Our scanner searches all assignments for ROM-like signals using the allowlist. For each assignment, the assigned value must not contain the keyword $wdata$. Otherwise, we flag a violation.

\textbf{CWE-1329}: Reliance on Component That is Not Updateable. 
This scanner checks all port connections for security-sensitive keywords from the allowlist. Then, if the connected signal is a constant value, we flag a violation.

\subsection{Bug Localization and Auto-fix}\label{sec:Auto-fix}

It can be time-consuming for a designer to debug a warning message, find the offending code, and write the fix. As part of our framework, we support bug localization and auto-fix suggestion generation. 

Bug localization is performed by marking the node (token) of each asset in the CST. Then, upon finding an offending pattern, we can look up the matching token location and provide the code location. Furthermore, we are able to update this token and its neighbours, allowing us to propose auto-fix suggestions. The tool provides an interactive patching method, but this is by no means a complete solution for automatic code repair, and is meant only as a starting point. 

We identify three types of auto-fixes: \texttt{add}, \texttt{replace}, and \texttt{remove}. Each type of fix involves the corresponding action its name would suggest (e.g., \texttt{remove} auto-fixes involve removing a section of code). Additionally, if we remove a particular symbol, we must also clean up any operators that were applied to that symbol. We perform cleanup by enforcing the following set of rules:

\begin{enumerate}
    \item Remove NOT operators with no operands, e.g., \verb|!()|
    \item Remove empty parentheses
    \item Remove back-to-back operators e.g., \verb|&& &&|. Keep the second operator and remove the first.
    \item Remove parentheses adjacent to operators, e.g., \verb|(&&|
\end{enumerate}

\autoref{tab:autofix_support} shows a summary of the auto-fix feature added to our tool. We identify the auto-fix type and the chosen operation to perform that auto-fix. Our implementation uses regular expressions to identify offending symbols and replace or delete them. 

\begin{table}[!ht]
    \small
    \centering
    \caption{\small Auto-fix Details For Supported CWEs}
    \label{tab:autofix_support}
    \begin{tabular}{|p{0.08\linewidth}|p{0.12\linewidth}|p{0.13\linewidth}|p{0.5\linewidth}|} 
    \hline
        \textbf{CWE} & \textbf{Auto-fix Support (Y/N)} & \textbf{Auto-fix type} & \textbf{Operation} \\ \hline
        1231 & Y & Remove & Remove debug signal and cleanup operators \\ \hline
        1232 & Y & Replace & Replace reset value with ‘1 (all ones) \\ \hline
        1234 & Y & Remove & Remove debug signal and cleanup operators \\ \hline
        1243 & Y & Replace & Replace ternary true value with ‘h0 (if ternary exists) \\ \hline
        1244 & Y & Replace & Replaces $||$ with \&\& \\ \hline
        1326 & Y & Remove & Remove entire assignment \\ \hline
    \end{tabular}
\end{table}

\section{Experimental Work}\label{sec:Experimental Work}

\subsection{Measuring Tool Performance}\label{sec:Measuring Tool Performance}

It is not clear from previous work \cite{ahmad_dont_2022} how to quantify the performance of our static analysis tools. Even though it is a binary classification problem, it is not easy to determine what constitutes a negative case---one would typically consider this a missed weakness, but in our case, we do not have knowledge of all weaknesses in the design. To expand on this, consider the well reputed benchmark we used earlier, the HACK@DAC21 SoC \cite{dessouky_hardfails:_2019} \cite{noauthor_hack_eventhackatdac21_2024}. Although there exists an informal bug list on the repository, \textbf{all relevant weaknesses are not entirely captured or localized}. Hence, we will assume that we can only access the true positive and false positive cases (which can be obtained through manually analyzing reported warnings from the tool). Therefore, our primary metric will be precision, which is defined as
\begin{equation}
    P = \frac{TP}{TP+FP}
\end{equation}

where $TP$ is the number of true positives and $FP$ is the number of false positives.

In addition, we will separate our binary classification evaluations for the asset identification component and the pattern matching component. That is, for asset identification, a true positive indicates we have correctly identified an asset; a false positive indicates we have incorrectly flagged a signal as an asset when it was not. In the case of our pattern checkers, a true positive indicates we have correctly identified a syntax pattern that, assuming it affects an asset, would constitute a potential weakness. A false positive indicates that, even assuming the syntax pattern affects an asset, would not constitute a potential weakness. 

As mentioned earlier, when scanning \textbf{natural} designs (designs without manually inserted bugs), the ground truth is typically unclear. Consequently, one might think that a \textbf{synthetic} design (a design with intentionally inserted bugs) would fix this issue. There are indeed synthetic datasets, such as those offered in related work~\cite{tarek_bugwhisperer_2025} \cite{bolcskei_encarsia_2025} \cite{noauthor_trust-huborg_nodate} \cite{cai_secrisben_2024} \cite{dessouky_hardfails:_2019}. However, we find in our analysis that these datasets may contain \textit{weaknesses} that are not necessarily the intended bug insertion (which we find for \cite{tarek_bugwhisperer_2025} in \autoref{sec:LLM-Generated Dataset}). Furthermore, synthetic datasets like \cite{bolcskei_encarsia_2025} are not compatible with our checker algorithms due to their gate-level netlist format, which removes the RTL patterns our algorithms search for.

As such, we set our sights on the HACK@EVENT SoCs~\cite{dessouky_hardfails:_2019}, since they contain the most relevant bugs and were studied in the previous work on static analysis \cite{ahmad_dont_2022}. However, as we discussed earlier, these datasets do not capture all potential weaknesses. Hence, we rely on manual analysis of the HACK@EVENT series SoCs \cite{dessouky_hardfails:_2019} to measure individual algorithmic performance.

We note that the code snippets from the CWE definitions on the MITRE site \cite{noauthor_cwe_nodate} often resuse code from the HACK@EVENT SoCs \cite{dessouky_hardfails:_2019}, making it non-distinct from our dataset of choice. 

\subsection{Analysis Methodology}\label{sec:Analysis Methodology}
With the discussion from \autoref{sec:Measuring Tool Performance} in mind, we provide our initial benchmarking results using the two SoCs from the HACK@EVENT series: HACK@DAC21 \cite{noauthor_hack_eventhackatdac21_2024} \cite{dessouky_hardfails:_2019} and HACK@DAC18 \cite{noauthor_hack_eventhackatdac18_2025} \cite{dessouky_hardfails:_2019}. In order to capture all potential weaknesses faithfully, we rely on manual analysis to label the code flagged by the tool. We identify violations as true positive (T) or false positive (F) (not allowing indeterminate (I) results, as in \cite{ahmad_dont_2022}) for both the asset identification aspect as well as the pattern checker aspect.

Our methodology for analysis is as follows: First, verify whether the signal was correctly identified as an asset. We define an asset as any signal that may plausibly relate to the security of the design. Mark this as T or F in the asset identification category. Then, completely separate to the first evaluation (assuming the signal is indeed an asset), verify if the CWE rule is violated, constituting a weakness. Mark this as T or F in the pattern checker category. Then, we can derive the overall classification. If both categories are T, then overall the classification is T. Otherwise, the overall classification is F. 

For example, consider the code in \autoref{fig:analysis_example_cwe1329_true}. Our scanner identifies the module port key\_i, which we classify as a security-relevant signal. The port value is indeed hard-coded, so it violates the rule for CWE-1329. Hence, we mark both categories as T. In contrast, we also identify the signal key\_hash\_bypass\_i as hard-coded. However, we do not classify this signal as security-relevant. Hence, we mark the asset identification as F but the pattern checker as T. 

\begin{figure}[!ht]
\centering
    \parbox{0.95\columnwidth}{
\begin{lstlisting}[language=verilog]
hmac hmac(
        .clk_i(clk_i),
        .rst_ni(rst_ni),
        .init_i(startHash),
        .key_i(256'h24e6fa2254c2ff632a41b3b4
               2e5c9b54f247b9a445a1ce488cfa23b384632154),
        .ikey_hash_i(ikey_hash_i), 
        .okey_hash_i(okey_hash_i), 
        .key_hash_bypass_i(1'h1),
        .message_i(pass_data),
        .hash_o(pass_hash),
        .ready_o(hmac_ready),
        .hash_valid_o(hashValid)   
    );
\end{lstlisting}
    }
    \caption{\small Code snippet from HACK@DAC21 \cite{dessouky_hardfails:_2019} \cite{noauthor_hack_eventhackatdac21_2024} highlighting a violation of CWE-1329} \label{fig:analysis_example_cwe1329_true}
\hfill%
\end{figure}

\subsection{Algorithmic Optimization Results}\label{sec:Algorithmic Optimization Results}

\begin{figure}[t]
    \centering
    \includegraphics[width=1\linewidth]{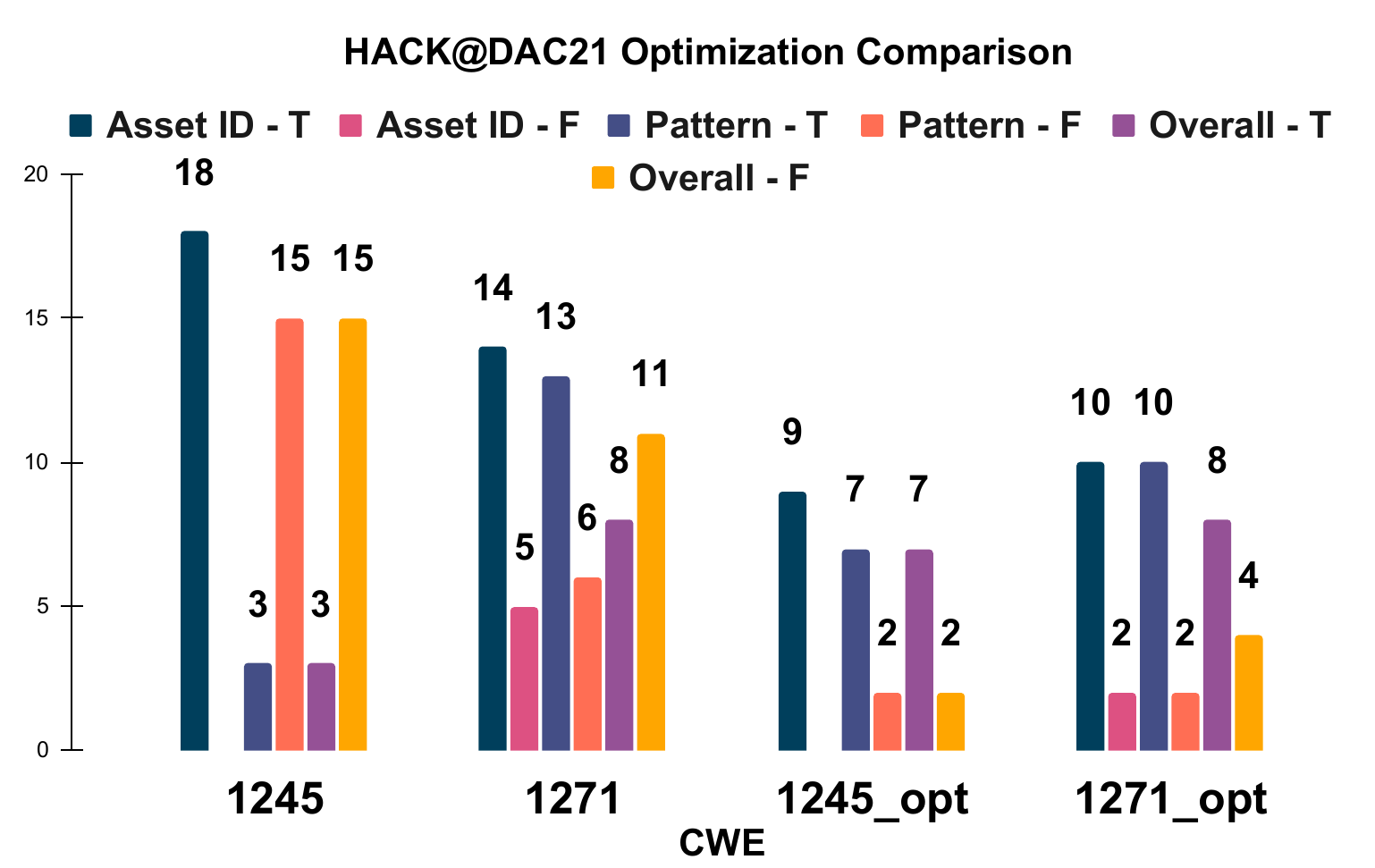}
    \caption{\small Comparison of the CWE-1245 and CWE-1271 algorithms with and without optimizations using the HACK@DAC21 \cite{dessouky_hardfails:_2019} \cite{noauthor_hack_eventhackatdac21_2024} dataset. After optimization, CWE-1245 sees an increase in overall precision from 0.167 to 0.778. For CWE-1271, this increases from 0.421 to 0.667. Compared to the results in \cite{ahmad_dont_2022}, the optimized CWE-1245 algorithm sees a 80\% decrease in false positives and a 75\% increase in true positives; the optimized CWE-1271 algorithm sees a 20\% decrease in false positives and 33.3\% increase in true positives.}
    \label{fig:hackatdac21_opt_comparison}
\end{figure}

\autoref{fig:hackatdac21_opt_comparison} shows our results analyzing the HACK@DAC21 SoC \cite{noauthor_hack_eventhackatdac21_2024} \cite{dessouky_hardfails:_2019} with and without any optimizations in the CWE-1245 and CWE-1271 algorithms. Before optimization, CWE-1245 suffered from a high false positive rate in its pattern checker, which was the bottleneck for the overall precision; after optimization, we improved the pattern checker, which improved overall precision. For CWE-1271, our pattern checker optimization helped to remove seven overall false positives. Note that our optimizations removed warnings, which changed the total counts. The remaining of the algorithms from CWEAT \cite{ahmad_dont_2022}, based on our initial analysis in \autoref{sec:Revisiting Prior Work}, did not warrant further optimization. Furthermore, due to our earlier work on the keyword list (\autoref{sec:Revisiting Prior Work}), we find that there are significantly less cases where the asset identification stage is the bottleneck for overall classification. 

\subsection{Benchmarking Results}\label{sec:Benchmarking Results}

We ran our first set of experiments on the HACK@DAC21 \cite{noauthor_hack_eventhackatdac21_2024} \cite{dessouky_hardfails:_2019} and HACK@DAC18 \cite{noauthor_hack_eventhackatdac18_2025} \cite{dessouky_hardfails:_2019} SoCs. For each design, we attempt to scan all Verilog and SystemVerilog files in the code base. When the tool encounters syntax errors while parsing the file, we skip that file and mark it as not scanned. We also ignore files that we manually identify as duplicates during our analysis, marking them as not scanned. All aforementioned algorithmic optimizations are enabled in this run. The results are summarized in \autoref{tab:violation_count_comparison}. The average runtime was 135.1 seconds, run on an Intel Core i5 processor with 16 GB of RAM, scanning an average of 483 files per run. We attribute the difference in runtime mainly to the different implementation of our tools. 

\autoref{tab:violation_count_comparison} compares the total violation counts of the algorithms in \cite{ahmad_dont_2022} with our implementation of these algorithms. \toolname{} takes longer to run, requiring an average of 135.1 seconds compared to the average time of 0.338 seconds in \cite{ahmad_dont_2022}. However, \toolname{} gives fewer total warnings, with a total of 43 warnings across both designs compared to the total of 119 in \cite{ahmad_dont_2022}. We note that our total file counts differ, indicating a difference in our file scanning methodologies. 

We observe that our checkers for CWE-1244 and CWE-1260 raise zero warnings. For both of these CWEs, the code examples are from HACK@DAC19 \cite{dessouky_hardfails:_2019} \cite{noauthor_hack_eventhackatdac19_2025}, which we do not examine because it does not provide a bug list. Perhaps the code examples for these CWEs are niche and our resulting heuristics are too selective. 

\begin{table*}[t]
    \small
    \centering
    \caption{\small Comparison of Total Violation Counts Between \toolname{} and CWEAT \cite{ahmad_dont_2022}}
    \label{tab:violation_count_comparison}
    \begin{tabular}{|p{0.12\linewidth}|p{0.2\linewidth}|p{0.2\linewidth}|p{0.2\linewidth}|p{0.2\linewidth}|}
    \hline
        \textbf{Design} & \textbf{HACK@DAC21 (Openpiton)} & \textbf{HACK@DAC21 (Openpiton)} & \textbf{HACK@DAC18 (Pulpissimo)} & \textbf{HACK@DAC18 (Pulpissimo)} \\ \hline
        \textbf{Tool} & \textbf{\toolname{} (Ours)} & \textbf{CWEAT} & \textbf{\toolname{} (Ours})\textbf{} & \textbf{CWEAT} \\ \hline
        Files Scanned & 480 & 328 & 486 & 409 \\ \hline
        File not Scanned & 6 & 112 & 7 & 116 \\ \hline
        Time (s) & 97.6 & 0.354 & 172.6 & 0.322 \\ \hline
        CWE-1234 & 2 & 2 & 0 & 0 \\ \hline
        CWE-1245 & 8 & 17 & 9 & 26 \\ \hline
        CWE-1262 & 1 & 10 & 0 & 10 \\ \hline
        CWE-1271 & 12 & 13 & 9 & 9 \\ \hline
        CWE-1280 & 2 & 12 & 0 & 20 \\ \hline
    \end{tabular}
\end{table*}

\begin{figure}[!ht]
    \centering
    \includegraphics[width=1\linewidth]{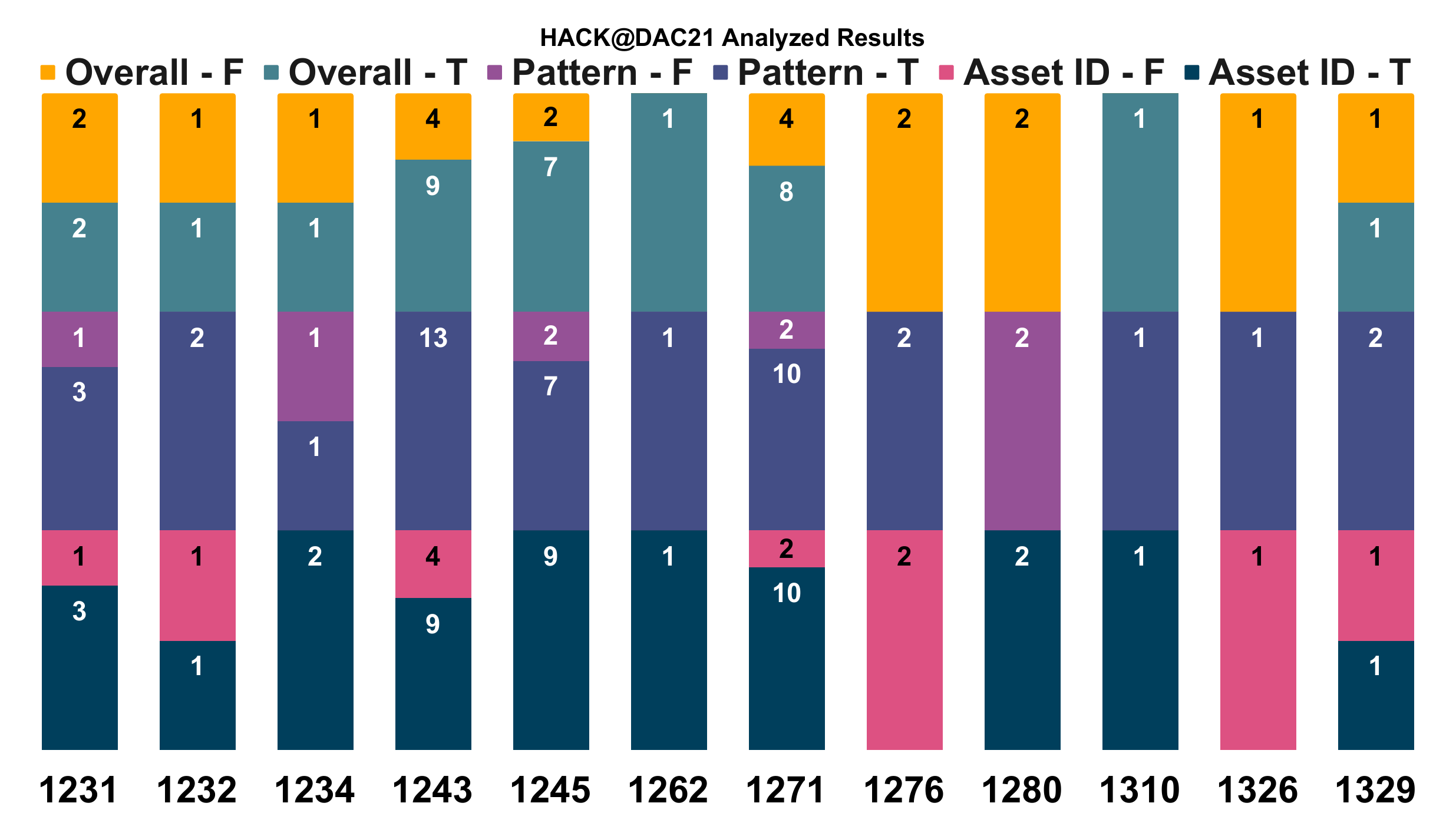}
    \caption{\small \toolname{} analyzed results for HACK@DAC21 \cite{dessouky_hardfails:_2019} \cite{noauthor_hack_eventhackatdac21_2024}. The horizontal axis shows each CWE for which we have implemented a scanner algorithm. The vertical axis represents the count of violations raised by \toolname{} for a given CWE scanner algorithm. CWE scanners that reported no warnings are not shown.}
    \label{fig:hackatdac21analysis}
\end{figure}

\begin{figure}[!ht]
    \centering
    \includegraphics[width=1\columnwidth]{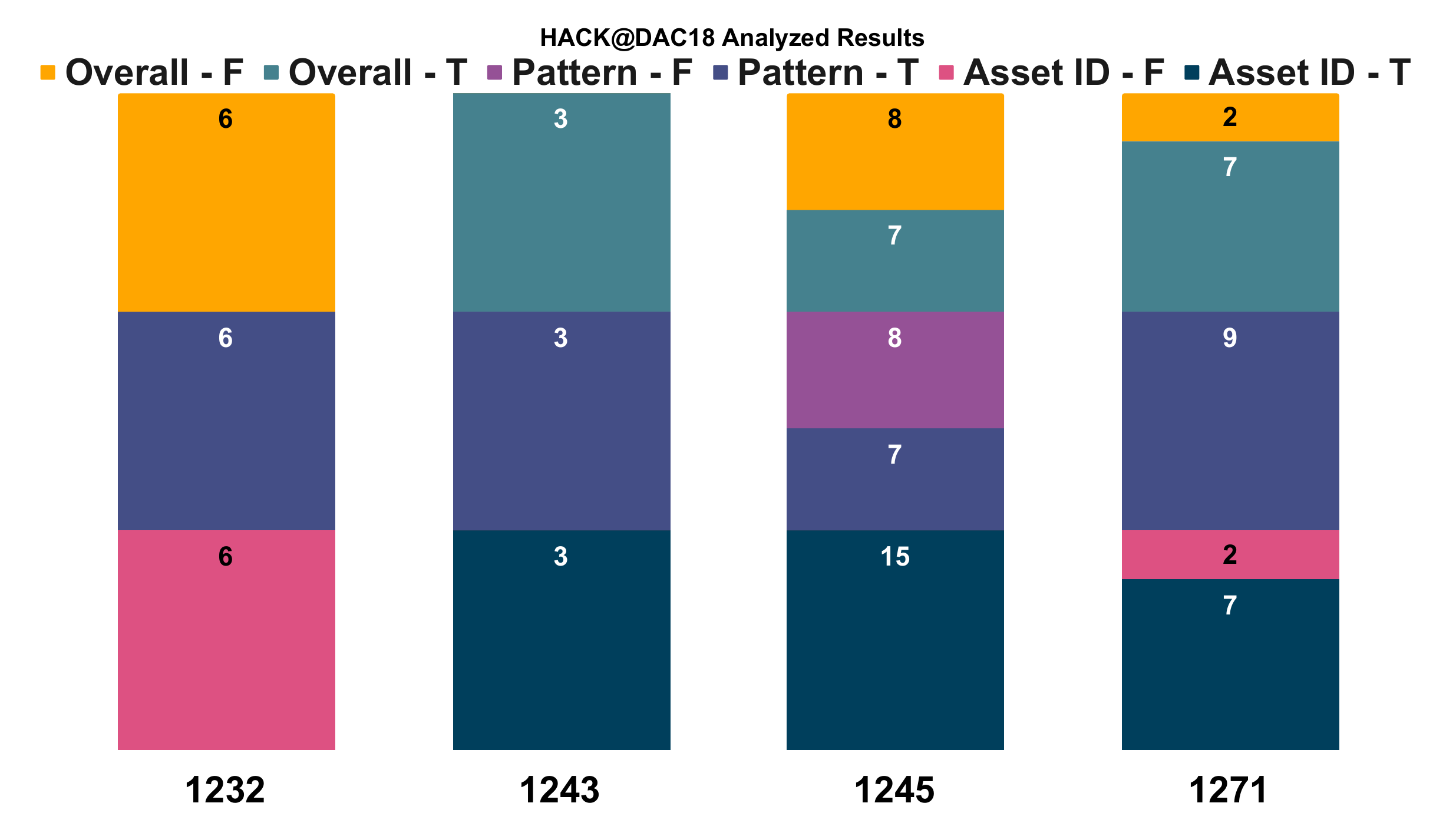}
    \caption{\small \toolname{} analyzed results for HACK@DAC18 \cite{dessouky_hardfails:_2019} \cite{noauthor_hack_eventhackatdac18_2025}. The horizontal axis shows each CWE for which we have implemented a scanner algorithm. The vertical axis represents the count of violations raised by \toolname{} for a given CWE scaner algorithm. CWE scanners that reported no warnings are not shown.}
    \label{fig:hackatdac18analysis}
\end{figure}

To provide a more useful measure of merit, we manually analyze the results, as summarized in \autoref{fig:hackatdac21analysis} and \autoref{fig:hackatdac18analysis}, using the methodology described in \autoref{sec:Analysis Methodology}. 

For HACK@DAC21, \toolname{} raises 51 errors. In our opinion, this is manageable for a designer to review. Of these 51 warnings, 31 were overall T and 20 were overall F. For HACK@DAC18, there were 33 total warnings, of which 17 were overall T and 16 were overall F. The overall precision was 0.608 in HACK@DAC21 and 0.515 in HACK@DAC18. Our asset identification precision was 0.765 for HACK@DAC21 and 0.758 for HACK@DAC18; our pattern precision was 0.843 in HACK@DAC21 and 0.758 for HACK@DAC18. Even though the two component precisions for asset identification and the pattern checkers were decent, they did not align their correctness simultaneously, reducing the overall precision. 

Although not exactly comparable due to variations in manual analysis, we provide the published results of CWEAT from Ahmad et al.'s paper~\cite{ahmad_dont_2022} in \autoref{fig:cweat_analysis} for reference. As a general trend, we see that CWEAT tends to produce considerable noise, i.e., warnings that are not T. Based on the analysis from \cite{ahmad_dont_2022}, CWEAT correctly raises a warning 17.5\% of the time for HACK@DAC21 and 6.2\% of the time for HACK@DAC18.

It is interesting to note that the overall precision for several of the scanners were mainly limited by incorrectness in asset identification. For example, CWE-1276, CWE-1326, and CWE-1329 were almost completely dominated by incorrect asset identification (but correct pattern identification). 

\begin{figure}[t]
    \centering
    \small
    \subfloat[HACK@DAC21]{
    \begin{tabular}{|c|c|c|c|}
    \hline
        \textbf{CWE} & \textbf{T} & \textbf{F} & \textbf{I} \\ \hline
        1234 & 0 & 0 & 2 \\ \hline
        1245 & 4 & 10 & 3 \\ \hline
        1262 & 1 & 6 & 2 \\ \hline
        1271 & 6 & 5 & 2 \\ \hline
        1280 & 0 & 12 & 0 \\ \hline
    \end{tabular}
    \label{fig:cweat_hackatdac21_analysis}}
    \subfloat[HACK@DAC18]{
    \begin{tabular}{|c|c|c|c|}
    \hline
        \textbf{CWE} & \textbf{T} & \textbf{F} & \textbf{I} \\ \hline
        1234 & 0 & 0 & 0 \\ \hline
        1245 & 4 & 21 & 1 \\ \hline
        1262 & 0 & 9 & 1 \\ \hline
        1271 & 0 & 1 & 8 \\ \hline
        1280 & 0 & 20 & 0 \\ \hline
    \end{tabular}
    \label{fig:cweat_hackatdac18_analysis}}
    
    \caption{\small Original CWEAT analyzed results from \cite{ahmad_dont_2022} for HACK@DAC21 \cite{dessouky_hardfails:_2019} \cite{noauthor_hack_eventhackatdac21_2024} and HACK@DAC18 \cite{dessouky_hardfails:_2019} \cite{noauthor_hack_eventhackatdac18_2025}.}
    \label{fig:cweat_analysis}
\end{figure}

\subsection{Investigating the BugWhisperer \cite{tarek_bugwhisperer_2025} Dataset}\label{sec:LLM-Generated Dataset}
Here, we use \toolname{} to explore a \textbf{synthetic} dataset that was generated using a large language model (LLM)~\cite{tarek_bugwhisperer_2025}. We entertained the idea of using the injected bug database from \cite{bolcskei_encarsia_2025}, but the tool generates netlists, which no longer contain the syntax patterns we are looking for. 

To parse the dataset from \cite{tarek_bugwhisperer_2025}, we convert each entry into a SystemVerilog file that contains the module definition. We then run \toolname{} on each generated file. The results of our scan are shown in \autoref{tab:bugwhisperer_counts}. The total runtime was 189.1 seconds, scanning a total of 3874 modules, 425 of which were ignored due to syntax errors. Given the sheer volume of violations, we opt to skip manual analysis.

The authors of \cite{tarek_bugwhisperer_2025} identify 13 CWEs for which the LLM targets its bug generation. Of these, we support scanners for CWE-1244, CWE-1245, CWE-1260, and CWE-1271. We do find violations within the dataset for CWE-1245, CWE-1260, and CWE-1271 as expected. However, we do not find any violations for CWE-1244, which reinforces the observation in \autoref{sec:Benchmarking Results} that this algorithm is, perhaps, too selective. Interestingly, we find violations for CWE-1231, CWE-1234, and CWE-1243, which were not part of the 13 intended CWEs injected by the LLM. 
We selected a small sample of the warnings from these three weaknesses and confirmed that there were indeed some overall true positives. 

This analysis revealed that the bugs within these categories almost completely reuse the same handful of module templates with little variation. Essentially, there is only one unique bug.

\begin{table}[t]
    \centering
    \small
    \caption{\small Total Violation Counts for LLM-generated Dataset}
    \label{tab:bugwhisperer_counts}
    \begin{tabular}{|c|c|c|c|}
    \hline
        \textbf{CWE} & \textbf{Violation Count} & \textbf{CWE} & \textbf{Violation Count}\\ \hline
        1231 & 14 & 1262 & 0 \\ \hline
        1232 & 0 & 1271 & 97 \\ \hline
        1234 & 37 & 1276 & 0 \\ \hline
        1243 & 1 & 1280 & 0 \\ \hline
        1244 & 0 & 1310 & 0 \\ \hline
        1245 & 541 & 1326 & 0 \\ \hline
        1260 & 6 & 1329 & 0 \\ \hline
    \end{tabular}
\end{table}

\subsection{Auto-fix Results}\label{sec:Auto-fix Results}
Using our two manually analyzed HACK@EVENT \cite{dessouky_hardfails:_2019} designs, we apply the auto-fix rules detailed in \autoref{sec:Auto-fix} to the applicable warnings. We declare an auto-fix T if there was indeed a security bug and it is fixed by the suggestion; we declare F otherwise. Note that in order for an auto-fix suggestion to be correct, we must have correctly identified the syntax pattern as well as the asset. Hence, the auto-fix classification will practically follow the overall classification. However, there is an exception in the case that we identify a syntax pattern that is a subset of a larger offending pattern. The overall classification might be T, since we have indeed identified a weakness, but the auto-fix might fail since it does not fix the overall weakness. 

For the HACK@EVENT \cite{dessouky_hardfails:_2019} designs, our auto-fix classification follows the overall classification exactly. 

For the dataset in \cite{tarek_bugwhisperer_2025}, we use our manual analysis of the 51 total violations flagged for CWE-1231 and CWE-1234. This time, we find 14 F auto-fixes for CWE-1231. For CWE-1234, we find 21 T and 16 F. For the 16 F cases in CWE-1234, we observed that the aforementioned exception resulted in an incorrect auto-fix despite a correct overall detection. 

Consider the auto-fix shown in \autoref{fig:autofix_cwe1234_bugwhisperer_incorrect}, where \toolname{} correctly issues a warning for CWE-1234, but the proposed auto-fix is not completely correct. In the comment accompanying this entry in the dataset, it states that the bug inserted in this module is the ability to access data from debug. Our auto-fix suggestion removes the debug\_mode signal, but does not remove the debug\_access signal, which we believe is incorrect. 

The detections for CWE-1231 were all overall F, which was due to incorrect asset identification. Hence, the modification to that signal in the suggested auto-fix is incorrect since it impacts the functionality of a non-security-relevant signal.

We observe that the auto-fix correctness is largely dependent on the overall precision of the detector. Hence, we conclude that this feature should be used to ease the manual burden of the designer, but still requires their judgment and supervision. 

\begin{figure}[t]
\centering
    \parbox{0.95\columnwidth}{
\begin{lstlisting}[language=verilog]
--- a/bugwhisperer/3081.sv 
+++ b/bugwhisperer/3081.sv @@ -34,6 +34,6 @@  
always @(posedge clk) begin 
-    if (state == READ && (debug_mode || !debug_access)) begin 
+    if (state == READ && (!debug_access)) begin          
read_data <= mem[addr];      end 
-    if (state == WRITE && (debug_mode || !debug_access)) begin 
+    if (state == WRITE && (!debug_access)) begin          
mem[addr] <= write_data; 
\end{lstlisting}
    }
    \caption{\small Example of an incorrect auto-fix on a true positive (T) warning for CWE-1234 from the BugWhisperer dataset \cite{tarek_bugwhisperer_2025}}
    \label{fig:autofix_cwe1234_bugwhisperer_incorrect}
\hfill%
\end{figure}

\section{Discussion}\label{sec:Discussion}

In this section, we discuss, in the context of lexical static analyis, RTL code pattern detection methods, the use of keyword lists, the efficacy of manual analysis, and datasets for benchmarking. 

\subsection{Code Pattern Detection}\label{sec:Code Pattern Detection}
Similar to \cite{ahmad_dont_2022}, we focus on detecting code patterns in the RTL. This approach is naturally difficult to generalize to all design types; we provide implementations for the common patterns seen in SoCs, which could be adapted for other classes of designs. E.g., using a preliminary study (using other verification methods or manual analysis), we could screen for the common patterns present within a given class of design. Additionally, our approach is unable to handle gate-level netlists and is not rigorous enough to replace other techniques. However, this is not our goal; this method should supplement other techniques by weeding out easily detected bugs before they propagate to later stages of verification.

\subsection{Keyword Lists}\label{sec:Keyword Lists}
A serious challenge is to identify specific functional structures (assets) within the design. To resolve this issue, we opt to use a keyword filtering strategy, which was used in previous work \cite{ahmad_dont_2022} \cite{deb_nath_toward_2025}. The keyword list may not be effective in other circumstances, so we provide support for the integration of emerging asset identification methods (e.g., LAsset \cite{hasan_lasset_2026}).

\subsection{Manual Analysis for Benchmarking}\label{sec:Manual Analysis for Benchmarking}
We rely on manual analysis to measure the improvements of our algorithmic optimizations, the efficacy of our auto-fix suggestions, and our relative performance compared to CWEAT \cite{ahmad_dont_2022}. Although we outline our classification methodology in \autoref{sec:Analysis Methodology}, there are bound to be errors and variations between different reviewers, creating some level of uncertainty in the analyzed results. Consequently, we avoid directly comparing our results with CWEAT and instead compare optimizations against our own baseline reimplementation. Furthermore, we make our full results available with this paper so that future work may reuse our analysis.

\subsection{Datasets for Static Analysis}\label{sec:Datasets for Static Analysis}
For consistent benchmarking, it would be ideal to have a dataset with all possible weaknesses. However, this is not available---and perhaps may be an infeasible task, given the ambiguity of the definition of a weakness. In hopes of improving this, however, we release our complete analysis in the accompanying repository to this paper. This work investigated the use of several datasets \cite{dessouky_hardfails:_2019,bolcskei_encarsia_2025,tarek_bugwhisperer_2025,noauthor_trust-huborg_nodate}. Of these, we were only able to use \cite{dessouky_hardfails:_2019} and, to some extent, \cite{tarek_bugwhisperer_2025}. We could not use \cite{bolcskei_encarsia_2025} since the netlist format was incompatible with our methodology, and we could not use \cite{noauthor_trust-huborg_nodate} since it focuses on hardware Trojans, which are not in the scope of this work.


Consequently, we evaluate our performance predominantly using the HACK@EVENT designs \cite{dessouky_hardfails:_2019} and manual analysis. Other works endorse this dataset for evaluating performance \cite{ahmad_dont_2022,tarek_socurellm_2025,saha_llm_2024,meng_rtl-contest_2022,rostami_fuzzerfly_2024}.

\subsection{Measuring Performance Beyond Precision}
One of the challenges with hardware security weakness detection is that negative cases are not well-defined, making it difficult to measure metrics such as recall. However, we provide a minimal case study in this regard for completeness. We consider the set of non-false-positive results that are available to us from CWEAT and determine the intersection and symmetric difference with \toolname's true positives. For the CWE-1271 checker, we find that \toolname{} discovers 2 TPs in HACK@DAC21 that were not found in CWEAT. Furthermore, \toolname{} does not miss any of the 6 TPs that CWEAT found. For CWE-1245, \toolname{} finds 3 TPs that CWEAT missed. \toolname{} also finds all of the TPs that CWEAT finds. For CWE-1234, we find both of the same cases CWEAT finds.

\section{Conclusion}\label{sec:Conclusion}

We present \toolname, a static analysis tool for security verification that combines lexical pattern checkers and asset identification to find potential weaknesses in early-stage RTL. \toolname{} expands on previous static analysis work, providing algorithmic optimizations and an expanded set of algorithms for more vulnerabilities. Additionally, unlike previous work, our implementation is able to identify the exact location in the RTL where the vulnerability occurred (bug localization) and provides automatic bug fix suggestions when applicable. To evaluate the relative performance of \toolname, we manually analyze a selection of synthetic datasets. We find that, compared to previous work, \toolname{} improved overall detection precision from 0.167 to 0.778 in the best case, among other improvements. We also analyze the effects that the two fundamental components, asset identification and pattern checking, have on overall detection performance. 

\appendices
\section{Artefacts}\label{sec:Artefacts}
Repository link: https://github.com/bryan-kwan/cweep

\bibliographystyle{IEEEtran}
\bibliography{acmart}

\end{document}